\documentstyle[12pt]{article}
\def\ba{\begin{eqnarray}}
\def\ea{\end{eqnarray}}

\def\t{\theta}

\def\lb{\label}
\def\be{\begin{equation}}
\def\ee{\end{equation}}

\begin{document}

\title{Toric hyperkahler manifolds with quaternionic Kahler bases
and supergravity solutions}
\author{Osvaldo P. Santillan \thanks{Bogoliubov Laboratory of Theoretical
Physics, JINR, 141 980 Dubna, Moscow Reg., Russia;
firenzecita@hotmail.com and osvaldo@thsun1.jinr.ru.}   and Andrei
G. Zorin \thanks{Faculty of Physics, MSU, Vorobjovy Gory, Moscow,
119899, Russia; zrn@aport.ru.} }
\date {}
\maketitle

\begin{abstract}

  In the present work some examples of toric hyperkahler metrics in eight dimensions
are constructed. First it is described how the Calderbank-Pedersen metrics arise
as a consequence of the Joyce description of selfdual structures in four dimensions,
the Jones-Tod correspondence and a result due to Tod and Przanowski. It is also shown
that any quaternionic Kahler metric with $T^2$ isometry is locally isometric to
a Calderbank-Pedersen one. The Swann construction of hyperkahler metrics in eight
dimensions is applied to them to find hyperkahler examples with
$U(1)\times U(1)$ isometry.
The connection with the Pedersen-Poon toric hyperkahler metrics
is explained and it is shown that there is a class of solutions of the generalized
monopole equation related to eigenfunctions
of certain linear equation. This hyperkahler examples are lifted to solutions
of the D=11 supergravity and type IIA and IIB backgrounds are found by use of dualities.
As before, all the description is achieved in terms of a single eigenfunction F. Some
explicit F are found, together with the Toda structure corresponding to the trajectories
of the Killing vectors of the Calderbank-Pedersen bases.

\end{abstract}

\section{Introduction}

   The relevance of hypergeometry in field theory
has been made manifest during the last twenty years. For example
the moduli space of magnetic monopoles \cite{Atiyo} or the moduli space of
Yang-Mills instantons in flat
space \cite{AHDM} are hyperkahler manifolds. The relation between hyperkahler spaces and
hypermultiplets of field theories in D=4 with
N=2 rigid supersymmetries has been pointed out in \cite{Gaume},
\cite{Hitchon} and \cite{Galicki} and it was
shown that when the supersymmetry is made local the hypermultiplets
couple to supergravity and the resulting target space is a quaternionic Kahler manifold \cite{Bagger}.
Many other modern applications of this subject to supersymmetric theories in D=4 can be found
in \cite{Fre}-\cite{deWit} and references therein.

   Quaternionic geometry is deeply related to gravity theories in different dimensions, and to superstring
and M-theories. This is because quaternionic Kahler metrics are
always Euclidean vacuum Einstein with cosmological constant $\Lambda$
and in the limit $\Lambda\rightarrow 0$ it is obtained an hyperkahler metric.
Four dimensional quaternionic metrics can be extended to examples of special holonomy \cite{Salamon},
which are internal spaces of supergravity theories preserving some amount of
supersymmetries. Moreover compact $G_2$-holonomy spaces with orbifold singularities
are believed to arise as a quotient of a conical hyperkahler manifold in D=8 by one of its isometries
\cite{Witten2}. Quaternionic manifolds also characterize the hypermultiplet geometry of classical and
perturbative moduli spaces of type II strings compactified on a Calabi-Yau manifold \cite{Rocco}.

   Hypergeometry is also an active tool in modern mathematics.
Quaternionic Kahler and hyperkahler spaces are D=$4n$ dimensional and constitute the special case
of the Berger list with holonomies included in $Sp(n)\times Sp(1)$ and $Sp(n)$ respectively \cite{Berger}.
Some of their properties has been investigated for instance in \cite{Wolf}- \cite{Proeyen} but they are not
completely classified at the present.

     One of the latest achievements in the subject is the
hyperkahler quotient, developed in \cite{Hitchon} and \cite{Kronheimer} and
providing a way to construct hyperkahler manifolds of a given
dimension taking the quotient of a higher dimensional hyperkahler one by certain group generating
triholomorphic isometries. A sort of inverse method is due to Swann \cite{Swann}
who shows how a quaternionic Kahler metric in D=$4n$ can be extended to a quaternionic Kahler and
hyperkahler examples in D=$4(n+1)$. The Swann construction was applied recently
to construct hyperkahler cones in \cite{Anguelova}, relevant in theories with N=2
rigid supersymmetries, and to construct certain scalar manifolds in M-theory
on a Calabi-Yau threefold in the vicinity of a flop transition \cite{mohap}.

     The present work is mainly focused in the construction of eight dimensional hyperkahler metrics
with two commuting $U(1)$ isometries (usually called toric hyperkahler) and the Swann extension is crucial to do
this. The reason is that the quaternionic Kahler metrics in $D=4$ with two
commuting $U(1)$ isometries has been locally completely classified
by Calderbank and Pedersen \cite{Pedersen} in terms of solutions F of a simple linear second
order equation, namely
$$
F_{\rho\rho}+F_{\eta\eta}=\frac{3F}{4\rho^2}.
$$
Such four dimensional metrics will be extended by the Swann construction to hyperkahler
ones in D=8 and it will be seen that the $T^2$ isometry is preserved in this extension, therefore the
result is toric hyperkahler. As in the Calderbank-Pedersen case, all the description is achieved in terms
of the linear equation given above, which make this picture very simple.

    There exist a physical motivation to construct toric hyperkahler examples. They arise
naturally in the M-theory context as solutions corresponding to multiple intersecting branes \cite{Gibbin},
but their range of applications is of course, not limited to this case. For instance, the moduli space of
scattering of well separated BPS monopoles or well separated dyons due to a $(p,q)$ string in a
D-3 brane are toric hyperkahler manifolds \cite{Manton}-\cite{Dyon}. The metric of the moduli
space of the $k=1$  $SU(n)$ periodic instantons (or calorons) has been shown to be toric
hyperkahler \cite{Caloron}. Applications related to intersections in domain walls can be found
in \cite{Tong} and \cite{Tong2} and in \cite{Portugues}, there has been studied solitons in a (2+1)-dimensional
sigma model with a toric hyperkahler target space preserving 1/2 of the
supersymmetries and their realization in M-theory.

    It will be of interest to compare the results presented here with the Pedersen-Poon description \cite{Poon} of
toric hyperkahler spaces, which is the most suitable for physical purposes. They statement is that for every
of such spaces there is a coordinate system in which they locally takes the generalized Gibbons-Hawking anzatz
$$
\overline{g}=U_{ij}dx^i\cdot dx^j+ U^{ij}(dt_i+A_i)(dt_j+A_j),
$$
in terms of solutions of the generalized monopole equations,
 namely, a pair $(U_{ij}, A_i)$ satisfying
$$
F_{x_{\mu}^i x_{\nu}^j}=\epsilon_{\mu\nu\lambda}\nabla_{x_{\lambda}^i}U_j,
$$
$$
\nabla_{x_{\lambda}^i}U_j=\nabla_{x_{\lambda}^j}U_i.
$$
It will be shown that this statement is true for the metrics presented here
and therefore it is again checked that they are toric hyperkahler. As a consequence
a family of solutions of the Pedersen-Poon monopole equation are found in term of
the eigenfunctions F presented above.

   To finish we recall that the Calderbank-Pedersen spaces are
related to Einstein-Weyl structures by the
Jones-Tod correspondence \cite{JonTod}, which states that for any
four dimensional selfdual space with at least one isometry
the space of the trajectories of the Killing vector is an Einstein-Weyl space
with a Toda structure defined over it. This statement applies for the Calderbank-Pedersen
spaces. Einstein-Weyl structures are described by the continuum limit of the Toda
equation \cite{Ward}
$$
(e^u)_{zz} + u_{xx}=0,
$$
and the Jones-Tod correspondence gives a map between this equation and the corresponding
for F. This fact is of interest because gives a recipe to find solutions of
a non-linear equation (the Toda one) by solving a linear one. This correspondence is
crucial to find the Einstein representatives among the conformal structures with selfdual
Weyl tensor with at least one isometry.

   The organization of the present work is as follows: in section 2 there
are described the Joyce spaces, which are the most general selfdual
conformal structures with two surface orthogonal commuting Killing vectors.
The underlying Toda structure of the Joyce spaces corresponding to the trajectories of
its Killing vectors is found in section 3 by use of the Jones-Tod correspondence.
In section 4 the quaternionic Kahler examples among them are found, that
is, the Calderbank-Pedersen metrics.
In section 5 the Swann construction is applied to them to find hyperkahler examples
with two commuting triholomorphic isometries.
The relation with the Pedersen-Poon metrics is explained in section 6 and it is find
a class of solutions of the Pedersen-Poon system in terms of an eigenfunction F.
Such form is the most suitable for physical purposes. As an application it is
shown in section 7 that the hyperkahler metrics of section 4 and 5
can be extended to different supergravity solutions by use of dualities.
In section 8 the Jones-Tod correspondence is used to generate some
implicit and explicit solutions of the equations mentioned above.
Section 9 contains the conclusions.

\section{Selfdual structures with two commuting isometries}

  In four dimensions to say that a manifold is quaternionic Kahler is equivalent
to say that is Einstein with selfdual Weyl tensor. For this reason in order
 to classify the toric quaternionic manifolds it is needed to
classify the selfdual structures with two commuting isometries in D=4.
Fortunately, there exists and complete classification of them made by Joyce
if the Killing vectors are surface orthogonal \cite{Joyce}. The demand of
$U(1)\times U(1)$ isometry and selfduality is very restrictive and in consequence
all the description is made in terms of solutions of a linear system of
differential equations. This section is intended to describe the Joyce classification
in the most simple way as possible, and the other two
are devoted to show which metrics among them are Einstein and thus toric quaternionic Kahler.

   It should be reminded that for an Euclidean space in D=4 the rotation group
$SO(4)$ is locally isomorphic to $SU(2)\times SU(2)$ and therefore the Weyl
tensor $W$ decomposes as $W=W_{+}\oplus W_{-}$ where the components $W_{\pm}$
corresponds to one of the $SU(2)$ groups. $W$ is by definition the conformally
invariant part of the Riemann tensor, this means that is unchanged under an scaling
$g\rightarrow \Omega^2 g$. A conformal structure $[g]$ is defined as the family of
metrics obtained from $g$ by conformal transformations. If $W_{-}=0$ for a given $g$ of $[g]$
then $g$ is called selfdual and, by conformal invariance, $[g]$
will be a selfdual structure.

   Let us focus in spaces $M$ with two commuting $U(1)$ isometries.
The manifolds in consideration are then of the form $M=N \times T^2$
where $N$ is a Riemann surface, and $T^2=U(1) \times U(1)$ is the two
dimensional flat torus. We will denote as $(\theta, \varphi)$ the periodic angles
parameterizing $T^2$. Consider an structure $[g]$ over $M$ with representatives
$g$ that locally takes the Gowdy form
\be\lb{form}
g= g_{ab}dx^a dx^b + g_{\alpha\beta}dx^{\alpha} dx^{\beta}.
\ee
The latin indices $a,b$ corresponds to vectors on $N$ and the greek indices
$\alpha, \beta$ to vectors on $T^2$. Both $g_{ab}$ and
$g_{\alpha\beta}$ are supposed to be independent of $x^{\alpha}=(\theta, \varphi)$.
It is seen that the Killing vectors are $\partial/\partial\theta$ and $\partial/\partial\varphi$
and the level surfaces of constant $\theta$ and $\varphi$ are orthogonal to both
Killing fields.

    By Gauss theorem there exists a local scale transformation $g \rightarrow \Omega^2 g$
which reduce (\ref{form}) to
\be\lb{form2}
g = d\rho^2+ d\eta^2 + \widetilde{g}_{\alpha\beta}dx^{\alpha} dx^{\beta}.
\ee
Because selfduality is a property of $[g]$ rather than $g$ there is not loss of
generality in consider the anzatz (\ref{form2}) instead of (\ref{form}).
Define the basis $(e_{1}, e_{2})$ such that
$$
\widetilde{g}_{\alpha\beta}dx^{\alpha} dx^{\beta}=e_{1}^2 + e_{2}^{2}.
$$
There exist a linear transformation $T$ connecting the basis $(\rho d\theta, \rho d\varphi)$
with $(e_{1}, e_{2})$, which we will write as
$$
T=\left(\begin{array}{cc}
  A_0 & A_1 \\
  B_0 & B_1
\end{array}\right),
$$
where $A_i$ and $B_i$ are certain functions of $(\rho, \eta)$. By calculating $T^{-1}$
it is seen that the angular part of $g$ can be expressed as
\be\lb{simplif}
\widetilde{g}_{\alpha\beta}dx^{\alpha} dx^{\beta}=\frac{(\rho A_0 d\theta - \rho B_0 d\varphi)^2
+ (\rho A_1 d\theta - \rho B_1 d\varphi)^2}{(A_{0} B_{1}-A_{1}B_{0})^2}.
\ee
The advantage of this form is that the selfduality condition is equivalent to a system of linear equations.
Imposing the condition $W_{-}=0$ for (\ref{form2}) gives the following proposition \cite{Joyce}:
\\

{\bf Proposition 1 }{\it Any selfdual $g$ with two commuting killing vectors $\partial/\partial\theta$
and $\partial/\partial\varphi$ over $M=N \times T^2$ is locally conformal to a selfdual metric $g_{j}$
of the form
\be\lb{joycemetric}
g_{j}=(A_{0} B_{1}-A_{1}B_{0})\frac{d\rho^2 + d\eta^2}{\rho^2}+
\frac{(A_0 d\theta - B_0 d\varphi)^2
+ (A_1 d\theta - B_1 d\varphi)^2}{(A_{0} B_{1}-A_{1}B_{0})},
\ee
where the functions $A^i$ satisfies
\be\lb{joyce1}
(A_0)_{\rho}+(A_1)_{\eta}=A_0/\rho,
\ee
\be\lb{joyce2}
(A_0)_{\eta}-(A_1)_{\rho}=0,
\ee
and the same equations holds for $B_i$.}
\\

Equations (\ref{joyce1}) and (\ref{joyce2}) are equivalent to the condition\
$W_{-}=0$. The Joyce metrics (\ref{joycemetric}) are obtained by
introducing (\ref{simplif}) in (\ref{form2}) and making a conformal rescaling with
a factor $(A_{0} B_{1}-A_{1}B_{0})/\rho^2$. Such form is more convenient in order to
find the Einstein metrics among the Joyce ones.
Therefore the problem to find toric selfdual structures in D=4 has been
reduced to solve a linear system for $A^i$ and independently for $B^i$.
The original proof of proposition 1 has been obtained in a rather
different way than here; it is based on a method discussed in Appendix B.

   It should be noted that (\ref{joyce2}) implies that
\be\lb{potencial1}
A_0=G_\rho;\;\;\;\;\;\; A_1=G_{\eta},
\ee
for certain potential function $G$. Then (\ref{joyce1}) implies that
$G_{\rho\rho} + G_{\eta\eta}=G_{\rho}/\rho$.
Conversely (\ref{joyce1}) implies that
\be\lb{potencial1}
A_0=-\rho V_\eta;\;\;\;\;\;\; A_1=\rho V_{\rho},
\ee
and (\ref{joyce2}) gives the Ward monopole equation \cite{Ward}
\be\lb{Wardy}
V_{\eta\eta}+\rho^{-1}(\rho V_{\rho})_{\rho}=0.
\ee
which has been proven to describe hyperkahler metrics with two commuting
isometries. The relations
\be\lb{shu}
G_\rho=-\rho V_{\eta}=A_0;\;\;\;\;\;\;\ G_{\eta}=\rho V_{\rho}=A_1.
\ee
constitute a Backlund transformation allowing to find a monopole V
starting with a known $G$ or viceversa.
The functions $B_i$ can be also expressed in terms of another
potential functions $G'$ and $V'$ satisfying the same equations than $V$
and $G$.

\section{The Toda structure}

   This section presents some results of key importance in order to recognize
the Einstein metrics among the Joyce ones. But before to present them in more detail
we should state some important properties about Einstein-Weyl structures.
We recall from Appendix A that a 3-dimensional Einstein-Weyl structure is an structure $[h]$
characterized by a representative $h$ of the form (\ref{conformal}) and a connection $D$ preserving $[h]$,
namely
\be\lb{Eisno}
h=e^{u}(dx^2+dy^2)+dz^2,\;\;\;\;\;D_{a} h_{bc}=\omega_{a}h_{bc}
\;\;\;\;\;\omega=-u_{z}dz.
\ee
The function $u$ satisfies the $SU(\infty)$ Toda equation
\be\lb{Todin}
(e^u)_{zz} + u_{yy} + u_{xx}=0.
\ee
With this result it is possible to enunciate the Jones-Tod correspondence \cite{JonTod}
contained in the following proposition:
\\

{\bf Proposition 2 }{\it a) Consider an Einstein-Weyl structure $[h]$ in
D=3 and a representative $h$. Then the four dimensional metric
\be\lb{Converse}
g = U h + \frac{(dt+ A)^2}{U}
\ee
is selfdual with one Killing vector $\partial_t$ if the pair of functions
$(U, A)$ satisfies the generalized monopole equation
\be\lb{nose}
dA= *_{h}(dU - U \omega).
\ee
The Hodge star $*_{h}$ is taken with respect to $h_{ij}$ and $\omega$ is defined in terms of the
Toda solution by the third (\ref{Eisno}).

      b) Conversely if a given $g$ is selfdual and has one conformal Killing vector
$K^a$ then a conformal transformation can be performed in order that $K^a$
becomes a Killing vector $\partial_t$ and there exists a system of
coordinates in which $g$ takes the form (\ref{Converse}), being $h$
a representative of an Einstein-Weyl structure. The factor $\omega$ will
be obtained in this case through (\ref{nose}).}
\\

 From formula (\ref{Eisno}) it is seen that
$$
dU - \omega U= U_x dx + U_y dy+ (U_z + u_z U)dz=U_x dx + U_y dy+
e^{-u}(e^u U)_z dz
$$
and (\ref{nose}) is then explicitly
$$
dA=*_{h}(dU - \omega U)= U_{x} dz \wedge dy + U_{y} dx \wedge dz +
(U e^{u})_{z} dy \wedge dx.
$$
Therefore the integrability condition for the existence of $A$ is
\be\lb{integra}
(U e^u)_{zz} + U_{yy} + U_{xx}=0.
\ee

  In other words the Jones-Tod result states that for every four dimensional
selfdual space with at least one isometry,
the space of trajectories of the Killing vector define an Einstein-Weyl
structure in 3-dimensions, and conversely every 3-dimensional Einstein-Weyl
structure is the space of trajectories of a Killing field of a four
dimensional selfdual space. This result applies for the Joyce spaces (\ref{joycemetric}) as
long as they have two isometries. The Jones-Tod correspondence has been originally obtained by use of
minitwistor theory. But the advantage to reduce the Joyce metrics to the form
(\ref{Converse}) is that the following theorem \cite{Przanowski}-\cite{Todo}
can be applied to find the Einstein representatives:
\\

{\bf Proposition 3 }{\it Any selfdual Einstein metric $g$
with one Killing vector in D=4 there exist
a system of coordinates $(x,y,z,t)$ for which takes the form
\be\lb{Prza}
g = \frac{1}{z^2}[U (e^{u}(dx^2 + dy^2) + dz^2) + \frac{1}{U}(dt+ A)^2].
\ee
The functions $(U, A, u)$ are independent of the variable $t$ and satisfies
\be\lb{prosa1}
(e^u)_{zz} + u_{yy} + u_{xx}=0,
\ee
\be\lb{prosa2}
dA=U_{x} dz \wedge dy + U_{y} dx \wedge dz + (U e^{u})_{z} dy \wedge dx,
\ee
\be\lb{prosa3}
U=2-zu_{z}.
\ee
Conversely, any solution of (\ref{prosa1}), (\ref{prosa1})
and (\ref{prosa3}) define by (\ref{Prza}) a selfdual Einstein metric.}
\\

  It is easily seen that if the condition (\ref{prosa3})
is relaxed then proposition 3 reduces to the proposition 2 up to an scaling by $1/z^2$.
Then (\ref{prosa3}) is the condition to be satisfied in order to have an Einstein metric. It is
sufficient because it can be checked that the integrability condition
$$
(U e^u)_{zz} + U_{yy} + U_{xx}=0,
$$
is always satisfied for $U=2-zu_z$. In other words,
every $SU(\infty)$ Toda solution define a selfdual metric
by (\ref{Prza}). Then the problem to find the Einstein metrics among the Joyce ones is to reduce
them to the form (\ref{Converse}) and then to apply (\ref{prosa3}). The result will be an extra
relation between the functions $A_i$ and $B_i$ and the resulting metrics will be toric quaternionic Kahler.

    The first task is to find a new coordinate system $(x,y,z,t)$ for the Joyce metrics (\ref{joycemetric})
defined in terms of the old one $(\rho,\eta, \theta,\varphi)$ for which they are expressed as
\be\lb{Piza}
g = [U (e^{u}(dx^2 + dy^2) + dz^2) + \frac{1}{U}(dt+ A)^2].
\ee
according to (\ref{Converse}). To do this it is needed to write (\ref{joycemetric}) as
\be\lb{joycemetric2}
g_{j}=\frac{A_0 B_1 - A_1 B_0}{\rho^2(A_0^2+A_1^2)}((A_0^2+A_1^2)(d\rho^2+d\eta^2)
+ \rho^2 d\varphi^2)+\frac{A_0^2+A_1^2}{A_0 B_1 - A_1 B_0}(d\theta-\frac{(A_0 B_0 + A_1 B_1)d\varphi}{A_0^2+A_1^2})^2
\ee
and is seen that after rescaling by $\rho$ and identifying $t=\theta$ that it takes the form
(\ref{Converse}) with a metric $h$ and a monopole $(U,A)$ given by
\be\lb{einsaxial}
h=(A_0^2+A_1^2)(d\rho^2+d\eta^2)+ \rho^2 d\varphi^2 ,\;\;\;\;\;\;U=\frac{A_1 B_0-A_0 B_1}{\rho(A_0^2 + A_1^2)},
\ee
\be\lb{reform}
A=-\frac{(A_0 B_0 + A_1 B_1)}{A_0^2 + A_1^2}d\varphi.
\ee
The factor $\omega$ can be calculated through $dA= *_{h}(dU - U \omega)$ and is
\be\lb{omega}
\omega=-\frac{2A_{0}}{\rho(A_{0}^2 + A_{1}^2)}dG;\;\;\;\;
dG=-\rho V_{\eta}d\rho + \rho V_{\eta}d\eta.
\ee

     The next problem to find the coordinates $(x,y,z)$ for
which (\ref{einsaxial}) takes the form (\ref{Eisno}). The relation
$\omega=-u_{z}dz$ and (\ref{omega}) suggests that $dz=dG$ and therefore
$G=z$ up to a translation. Indeed, the other possible differential constructed with $V$ is
$$
dV= V_{\rho}d\rho + V_{\eta}d\eta,
$$
and it can be easily checked that
$$
dG^2 + \rho^2 dV^2 = \rho^2 (V^2_{\eta}+ V^2_{\rho})(d\rho^2 + d\eta^2)=
(A_0^2+A_1^2)(d\rho^2+d\eta^2),
$$
where in the last step formula (\ref{potencial1}) has been used.
From the last expression is seen that (\ref{einsaxial}) is
\be\lb{einsaxial2}
h=\rho^2(dV^2 + d\varphi^2) + dG^2.
\ee
Comparision between (\ref{einsaxial2}) and  (\ref{Eisno}) shows that
a solution $u(x,z)$ of the continuum Toda equation is defined
by the identifications
\be\lb{ensartado}
e^{u}=\rho^2,\;\;\;\;\;\; x= V,\;\;\;\;\;\;y=\varphi,\;\;\;\; z=G.
\ee
The solution $u$ is independent of y is due to the presence of the
other isometry, which is also a symmetry of $h$. Formula (\ref{ensartado})
defines the coordinate system that we were looking for.

    At first sight (\ref{ensartado}) relates the solutions of the
axially symmetric Toda equation with two solutions V and G of
two different linear differential equations. But they are related by
a Backlund transformation and it can be directly checked that if $V$
is a Ward monopole, then $W$ such that $W_{\eta}=V$ is also a
Ward monopole and it follows that $G=\rho W_{\rho}$.
Inserting the expressions in terms of $W$ in (\ref{ensartado})
and changing the notation replacing $W$ by $V$ by convenience
gives the following proposition \cite{Ward}:
\\

{\bf Proposition 4 }{\it Any solution $V$ of the equation
$V_{\eta\eta}+\rho^{-1}(\rho V_{\rho})_{\rho}=0$
defines locally the coordinate system $(x, z)$
\be\lb{Wardchan}
x=V_{\eta},\;\;\;\;\;\; z = \rho V_{\rho},
\ee
in terms of $(\rho, \eta)$ and conversely (\ref{Wardchan}) defines implicitly $(\rho, \eta)$
as functions of $(x,t)$. Then the function $u(x,z)=log(\rho^2)$
is a solution of the axially symmetric Toda equation
\be\lb{axialtoda}
(e^u)_{zz} + u_{xx}=0.
\ee
This procedure can be inverted in order to find a Ward monopole $V$
starting with a given Toda solution.}
\\

Proposition 4 gives a method to find solutions of a non linear equation
(the continuum Toda one) starting with a solution of a linear one (the
Ward equation). But it is difficult in practice to find explicit solutions
of (\ref{axialtoda}) and usually proposition 4 gives implicit solutions.

   An important detail is that the Toda structure (\ref{einsaxial}) and the Toda solution $u$
are completely determined just in terms of $A_i$. Only the monopole
$(U, A, \omega)$ depends on both  $A_i$ and $B_i$, which are not related
in any way.

\section{Quaternionic-Kahler metrics with $U(1) \times U(1)$ isometry}

   It is of special interest to determine which $g$ among the Joyce metrics
(\ref{joycemetric}) are Einstein; in four dimensions selfdual
Einstein spaces are quaternionic-Kahler \cite{Ishihara}.
This will be performed applying the Einstein condition (\ref{prosa3}) to (\ref{joycemetric})
and the result is the Calderbank and Pedersen metrics \cite{Pedersen}.

    However it has been shown in the previous section that (\ref{joycemetric}),
(\ref{joyce1}) and (\ref{joyce2}) describe all the toric selfdual metrics with surface
orthogonal Killing vectors, but there are examples that admit $T^2$ actions
for which surface orthogonality do not hold, even locally
(see \cite{Joyce} pag. 534). Nevertheless the Killing vectors of a
selfdual metric with $U(1)\times U(1)$ isometry are surface orthogonal if it is Einstein \cite{Pedersen} and
this implies that the Calderbank-Pedersen metrics are the most general toric
quaternionic-Kahler ones. This statement do not hold in the hyperkahler limit,
in which the scalar curvature tends to zero.

    For Joyce spaces the relation $\omega=-u_z dz$ and (\ref{omega})
gives
$$
u_z=\frac{A_0}{\rho(A_0^2+A_1^2)},\;\;\;\;\;\;
2-zu_z=\frac{\rho(A_0^2+A_1^2)-G A_0}{\rho(A_0^2+A_1^2)}.
$$
Then the insertion of the expression for $U$ (\ref{reform}) in terms of
$A_i$ and $B_i$ into the Einstein condition $U=2-zu_{z}$ gives
$$
A_1 B_0-A_0 B_1=\rho(A_0^2+A_1^2)-G A_0.
$$
Thus $B_0=\rho A_1 + \xi_0$ and $B_1=G-\rho A_0 + \xi_1$
with $A_1 \xi_0 = A_0 \xi_1$. The functions
$\xi_i$ are determined by asking $B_i$ to satisfy the Joyce system
(\ref{joyce1}) and (\ref{joyce2}), the result is
$\xi_0=-\eta A_0$ and $\xi_1=-\eta A_1$. Therefore the metric $g_{j}/\rho z^2$
is Einstein if and only if
\be\lb{Caldon1}
A_0=G_\rho;\;\;\;\;\;\; A_1=G_{\eta}
\ee
\be\lb{Caldon2}
B_0=\eta G_{\rho}-\rho G_{\eta};\;\;\;\;\;\; B_{1}=\rho G_{\rho} + \eta G_{\eta}-G,
\ee
which is the Calderbank-Pedersen solution. Defining $G=\sqrt{\rho}F$
it follows that F satisfies
$$
F_{\rho\rho} + F_{\eta\eta}=\frac{3F}{4\rho^2}.
$$
Then inserting (\ref{Caldon1}) and (\ref{Caldon2})
expressed in terms of F into $g_{j}/\rho z^2$ and making the
identification $z=G$ gives the following
proposition \cite{Pedersen}:
\\

{\bf Proposition 5 }{\it For any Einstein-metric with
selfdual Weyl tensor and nonzero scalar curvature possessing two linearly
independent commuting Killing fields there exists a coordinate system in which
the metric $g$ has locally the form
$$
ds^2=\frac{F^2-4\rho^2(F^2_{\rho}+F^2_{\eta})}{4F^2}\frac{d\rho^2+d\eta^2}{\rho^2}
$$
\be\lb{metric}
+\frac{[(F-2\rho F_{\rho})\alpha-2\rho F_{\eta}\beta]^2+[(F+2\rho F_{\rho})\beta -
2\rho F_{\eta}\alpha ]^2}{F^2[F^2-4\rho^2(F^2_{\rho}+F^2_{\eta})]},
\ee
where $\alpha=\sqrt{\rho}d\theta$ and $\beta=(d\varphi+\eta d\theta)/\sqrt{\rho}$
and $F(\rho, \eta)$ is a solution of the equation
\be\lb{backly}
 F_{\rho\rho} + F_{\eta\eta}=\frac{3F}{4\rho^2}.
\ee

on some open subset of the half-space $\rho>0$.
On the open set defined by $F^2 > 4\rho^2(F^2_{\rho}+F^2_{\eta})$ the metric $g$ has positive scalar curvature,
whereas $F^2 < 4\rho^2(F^2_{\rho}+F^2_{\eta})$ -$g$ is selfdual with negative
scalar curvature.}
\\

The Einstein condition $R_{ij}=\kappa g_{ij}$ is not invariant under scale
transformations, so Proposition 5 gives all the quaternionic-Kahler metrics
with $T^2$ isometry up to a constant multiple. The problem to find them
is reduced to find an F satisfying the linear equation (\ref{backly}), that is, an eigenfunction
of the hyperbolic laplacian with eigenvalue $3/4$.

  The equation for $V$ (\ref{Wardy})  and has solutions of the form
$$
V_1(\rho,\eta)=W(\eta, i\rho) + c.c,\;\;\;V_2(\rho,\eta)=W(i\eta, \rho) + c.c
$$
\be\lb{intexp}
W(\eta, \rho)=\frac{1}{2\pi}\int^{2\pi}_0 H(\rho sen(\theta)+ \eta)d\theta
\ee
where $H(z)$ is an arbitrary  function of one variable \cite{Ward}. The
Backlund relations (\ref{shu}) define $V$ in terms of $G$, and consequently in terms of $F$,
and viceversa. For instance, non trivial eigenfunctions F can be constructed
selecting an arbitrary $H(z)$, performing the integration (\ref{intexp}) and finding $G$
through (\ref{shu}), then $F=G/\sqrt{\rho}$. In the same way an axially symmetric
Toda solution $u$ can be constructed starting with an arbitrary $H(z)$ by using proposition
3.

\section{Toric hyperkahler geometry in eight dimensions}

    Four dimensional quaternionic-Kahler manifolds can be used as
base spaces to construct $G_2$ holonomy manifolds and 8 dimensional
hyperkahler ones, by use of the Bryant-Salamon \cite{Salamon} and Swann \cite{Swann}
constructions respectively. Both types of manifolds can be extended to different supergravity
solutions by use of dualities. The aim of the following two sections is to
construct the hyperkahler metrics corresponding to the class (\ref{metric}) by means to
the Swann extension. As it will be clear, the two $U(1)$ isometries
of the Calderbank-Pedersen metrics are extended to the resulting hyperkahler ones
and are triholomorphic, which means that preserve the complex structures
defined over them. The presence of the Killing vectors is of importance when dealing with compactification
because II supergravity backgrounds can be found starting
with 11 supergravity solutions by reduction along the isometries. This point
will be discussed in section 7 in more detail.

\subsection{Properties of quaternionic Kahler manifolds}
    Before to present the Swann construction it is convenient to review
certain properties of quaternionic manifolds \cite{Ishihara}. Consider a Riemanian
space $M$ of real dimension $4n$ endowed with a metric
$$
ds^2=g_{\mu\nu}(x)dx^{\mu}dx^{\nu}
$$
and a set of three almost complex structures $J^i$ with $i=1,2,3$ satisfying the quaternionic algebra
\be\lb{almcomp}
J^{i} \cdot J^{j}=-\delta_{ij}+\epsilon_{ijk}J^{k},
\ee
and for which the metric $g$ satisfies $g(J^iX,J^iY)=g(X,Y)$ for any X,Y in $T_x M$.
A metric for which the last condition holds is known as quaternionic hermitian and it
follows that $J^{i}_{\beta\alpha}=-J^{i}_{\alpha\beta}$. Any combination $C$ of the form
$$
C=a^i J^i, \;\;\;\; a^ia^i=1
$$
will be an almost complex structure too, so $M$ has a family of almost complex structures
parameterized by the space $S^2$ of unit imaginary quaternions. From the three almost complex structures
(\ref{almcomp}) one can define an $SU(2)$ "gauge field"
\be\lb{onefor}
\widetilde{A}^{i}_{\mu}=\omega^{mn}_{\mu}J^{i}_{mn},
\ee
and consequently an $SU(2)$-curvature
$$
F^i=d\widetilde{A}^i+\epsilon_{ijk}\widetilde{A}^j \wedge \widetilde{A}^k,
$$
where $\omega^{mn}$ is the antiselfdual part of the spin connection of $g$. Also it is possible to
generalize the Kahler form corresponding to complex manifolds to an hyperkahler triplet
$\Omega_i$ defined by
\be\lb{triplet}
\Omega^{i}=e^{m}\wedge J^{i}_{mn}e^{n}.
\ee
Then the manifold $M$ is quaternionic Kahler if
\be\lb{rela}
F^i=\kappa \Omega^i,
\ee
holds, being $\kappa$ the scalar curvature of $M$.

    If (\ref{rela}) is satisfied then the usual Bianchi identities
of gauge theories implies that
\be\lb{rela2}
\nabla_{\alpha}\Omega^{i}=d\Omega^{i}+\epsilon_{ijk}\widetilde{A}^j \wedge \Omega^{k}=0.
\ee
The relation (\ref{rela2}) shows that the hyperkahler form of every quaternionic Kahler space is covariantly
closed with respect to the connection $\widetilde{A}^i$. In the hyperkahler limit $\kappa\rightarrow 0$
and (\ref{rela}) shows that $A$ is a pure gauge and can be reduced to zero. Then
\be\lb{closed}
d\Omega^i=0,
\ee
that is the hyperkahler triplet of an hyperkahler manifold is closed.

     It can be shown \cite{Ishihara} that any quaternionic metric is an Einstein space with curvature $\kappa$ and
$$
R_{mn}=3\kappa g_{mn}.
$$
In four dimensions a quaternionic-Kahler metric is an Einstein metric with selfdual Weyl
tensor. The holonomy $H\subseteq Sp(1)\times Sp(n)$. In the hyperkahler limit
it is Ricci-flat and the holonomy is reduced to $H \subseteq Sp(n)$.

    In four dimensions we can select a selfdual complex structure ($J^i_{ab}=-\epsilon_{abcd}J^i_{cd}/2$)
and the components of $\Omega^i$ and the $SU(2)$ gauge field $A_{\mu}$ will be given explicitly by
\be\lb{hyperkaliente}
\Omega^{1}=e^{0}\wedge e^{3}- e^{1}\wedge e^{2},\;\;\;
\Omega^{2}=e^{0}\wedge e^{2}+ e^{3}\wedge e^{1},\;\;\;
\Omega^{3}=-e^{0}\wedge e^{1}+ e^{2}\wedge e^{3},
\ee
\be\lb{gaugen}
A^{1}=\omega_{\mu}^{03}- \omega_{\mu}^{12},\;\;\;
A^{2}=\omega_{\mu}^{02}+ \omega_{\mu}^{31},\;\;\;
A^{3}=-\omega_{\mu}^{01}+ \omega_{\mu}^{23}.
\ee

\subsection{The Swann extension}

   In order to construct toric hyperkahler metrics in eight dimensions it is convenient
to introduce the quaternionic notation used in $SU(2)$ gauge theory. A metric in D=4 will
be written as $g=e \overline{e}=|e|^2$ where the quaternionic valued einbein $e$ is $e=e_0 + e_i J^i$
and $\overline{e}$ is its quaternionic conjugated. In general for two pure quaternionic 1-form
$$
\mu=\mu_{0}+ \mu_{1}J^{1}+\mu_{2}J^{2}+\mu_{3}J^{3},\;\;\;\;\;
\nu=\nu_{0}+ \nu_{1}J^{1}+\nu_{2}J^{2}+\nu_{3}J^{3}
$$
the quaternionic wedge product is defined as
\be\lb{quaternionil}
\mu \wedge \nu=(\mu_0\wedge\nu_1-\mu_2\wedge\nu_3)J^1
+(\mu_0\wedge\nu_2-\mu_3\wedge\nu_1)J^2+(\mu_0\wedge\nu_3-\mu_1\wedge\nu_2)J^3.
\ee
$$
+\mu_0\wedge\nu_0 + \mu_1\wedge\nu_1+ \mu_2\wedge\nu_2 + \mu_3\wedge\nu_3 ,
$$
and in particular
\be\lb{quaternioni}
\overline{\mu} \wedge \mu=(\mu_0\wedge\mu_1-\mu_2\wedge\mu_3)J^1
+(\mu_0\wedge\mu_2-\mu_3\wedge\mu_1)J^2+(\mu_0\wedge\mu_3-\mu_1\wedge\mu_2)J^3
\ee
pure quaternionic components. Using (\ref{quaternioni}) the formulas
(\ref{hyperkaliente}) and (\ref{gaugen}) can be expressed more compactly as
\be\lb{quaterom}
\Omega=e\wedge \overline{e},\;\;\;\; A=A^i J^i.
\ee
Formula (\ref{quaterom}) can be easily generalized to higher dimensions, for instance,
a metric $g$ in eight dimensions can be expressed as $g=e_1\overline{e}_1+ e_2\overline{e}_2$
with two quaternion einbeins $e_1$ and $e_2$ and then
\be\lb{quaterom2}
\Omega=e_1\wedge \overline{e}_1+ e_2\wedge \overline{e}_2.
\ee

The quaternionic expression of the relations (\ref{rela2}) and (\ref{rela}) in D=$4n$
is expressed with the help of this notation as
\be\lb{gaugo}
d\Omega-\widetilde{A} \wedge \Omega + \Omega\wedge \widetilde{A}=0,\;\;\;
d\widetilde{A}-\widetilde{A} \wedge \widetilde{A} +\kappa\Omega=0.
\ee
It is important to present the Swann construction to express $\Omega$ and $d\Omega$
entirely in terms of $A$ and its derivatives. This is easily achieved introducing
the second (\ref{gaugo}) into the first to give
\be\lb{gaugo2}
\kappa d\Omega+ \widetilde{A}\wedge d\widetilde{A} - d\widetilde{A} \wedge \widetilde{A},\;\;\;
d\widetilde{A}-\widetilde{A} \wedge \widetilde{A} +\kappa\Omega=0.
\ee

Formulas (\ref{gaugo2}) are the desired result. They are very useful in order to extend
a quaternionic Kahler metric $g=e\overline{e}$ in $D=4$ with local coordinates $(x_1, .., x_4)$
to an hyperkahler one $\overline{g}$ in $D=8$ parameterized by $(x_1, .., x_4, q)$,
being $q= q_0 + q_i J^i$ a quaternionic coordinate.
To achieve this task first it should be noted that the quaternionic form $\Xi$
given by
$$
\Xi=dq \wedge (\kappa\Omega+ d\widetilde{A} - \widetilde{A} \wedge \widetilde{A})\overline{q}
+ q (\kappa\Omega+ d\widetilde{A} - \widetilde{A} \wedge \widetilde{A})\wedge d\overline{q}
$$
$$
+ q(\kappa d\Omega+ \widetilde{A}\wedge d\widetilde{A} - d\widetilde{A} \wedge \widetilde{A})\overline{q},
$$
is identically zero by (\ref{gaugo2}). By another side it is possible to express $\Xi$ as a differential $\Xi=d\Phi$,
where
\be\lb{quakal}
\Phi=\kappa q\Omega \overline{q}+ (dq + \widetilde{A} q)\wedge \overline{(dq + \widetilde{A} q)}.
\ee
The condition $\Xi=0$ then implies that $\Phi$ is closed. By
(\ref{quaternioni}) it follows that $\Phi$ is a pure quaternion and therefore
is a candidate to be the hyperkahler triplet of an hyperkahler metric $\overline{g}$.
From (\ref{quaterom2}) it is seen that the einbein of $\overline{g}$ should be $e_1=q e$
and $e_2= dq + q\widetilde{A}$. Then the hyperkahler metric
$\overline{g}=e_1\overline{e}_1+ e_2\overline{e}_2$ with (\ref{quakal})
as hyperkahler form is
$$
\overline{g}=\kappa|q|^2g+|dq+q\widetilde{A}|^2.
$$
Therefore we have obtained the Swann theorem in D=8, namely \cite{Swann}:
\\

{\bf Proposition 6 }{\it If a four dimensional quaternionic Kahler metric $g$
is given, then the eight dimensional metric
\be\lb{Swannmetric}
\overline{g}=\kappa|q|^2g+|dq+q\widetilde{A}|^2
\ee
is hyperkahler. The coordinate $q= q_{0}+ q_{1}J^{1} + q_{2}J^{2} + q_{3}J^{3}$
takes quaternionic values and the "SU(2) gauge field" $\widetilde{A}$ is defined by (\ref{onefor})}
\\

Proposition 6 provides an extension to four to eight dimensions but its converse
is not necessarily true, that is, not every eight dimensional
hyperkahler metric can be expressed as (\ref{Swannmetric}) with a quaternionic
Kahler base $g$. A counterexample will be given in the next section.

    Clearly proposition 4 applies to the Calderbank-Pedersen metrics (\ref{metric}) and, by use of
(\ref{Swannmetric}), it follows immediately a family of toric hyperkahler spaces for every solution of the equation
(\ref{backly}). In order to find them one should calculate the expressions (\ref{hyperkaliente}) and
(\ref{gaugen}) corresponding to (\ref{metric}), the result is \cite{Pedersen}
$$
\Omega^1=\frac{1}{F^2}(F^2-4\rho^2(F^2_{\rho} + F^2_{\eta}))(\frac{d\rho\wedge d\eta}{\rho^2}+
\alpha\wedge\beta),\;\;\;
\Omega^2=\frac{1}{F^2}(\rho F_{\eta}\beta +(\rho F_{\rho}-\frac{1}{2}F)\alpha )\wedge \frac{d\rho}{\rho},
$$
\be\lb{hyperform}
\Omega^3=\frac{1}{F^2}(\rho F_{\eta}\alpha-(\rho F_{\rho}+\frac{1}{2}F)\beta )\wedge \frac{d\eta}{\rho},
\ee
\be\lb{threefor}
\widetilde{A}^1=\frac{1}{F}((\frac{1}{2}F
+\rho F_{\rho})\frac{d\eta}{\rho}-\rho F_{\eta}\frac{d\eta}{\rho}),\;\;\;\;
\widetilde{A}^2=-\frac{\alpha}{F},\;\;\;\;\; \widetilde{A}^3=\frac{\beta}{F}.
\ee
The explicit form of the resulting $\overline{g}$ is
\be\lb{explico}
\overline{g}=\kappa|q|^2g_{cp}+(dq_o-q_i\widetilde{A}^i)^2
+(dq_i+ q_0\widetilde{A}^i + \epsilon_{ijk}q_k\widetilde{A}^j)^2
\ee
where $g_{cp}$ is (\ref{metric}) and $\widetilde{A}^i$ is given by (\ref{threefor}). Also from
(\ref{threefor}) and (\ref{hyperform}) follows an explicit expression for (\ref{quakal})
A new coordinate system for this metrics, more suitable for physical applications,
will be found in the next section.

\section{Connection with the Pedersen-Poon metrics}

    If the Calderbank-Pedersen metrics (\ref{metric}) are used as base spaces in the Swann
construction (\ref{Swannmetric}) the resulting metrics are (\ref{explico}) and the
two isometries corresponding to the Killing vectors $\partial/\partial\theta$ and
$\partial/\partial\varphi$ are preserved in this extension
and are triholomorphic, which means that
$$
\pounds_{\frac{\partial}{\partial\varphi}} J^i=0,\;\;\;\pounds_{\frac{\partial}{\partial\theta}} J^i=0.
$$
Therefore the result is a toric hyperkahler metric in eight dimensions.
But the hyperkahler metrics in D=$4n$ with $n$ commuting triholomorphic $U(1)$
isometries has been completely classified locally by Pedersen and Poon in terms of the
generalized Gibbons-Hawking anzatz. Their eight dimensional statement is \cite{Poon}:
\\

{\bf Proposition 7 }{\it For any hyperkahler metric in $D=8$ with two commuting triholomorphic
U(1) isometries there exists a coordinate system in which takes the form
\be\lb{gengibbhawk}
\overline{g}=U_{ij}dx^i\cdot dx^j+ U^{ij}(dt_i+A_i)(dt_j+A_j),
\ee
where $(U_{ij}, A_i)$ are solutions of the generalized monopole equation
$$
F_{x_{\mu}^i x_{\nu}^j}=\epsilon_{\mu\nu\lambda}\nabla_{x_{\lambda}^i}U_j,
$$
\be\lb{genmonop}
\nabla_{x_{\lambda}^i}U_j=\nabla_{x_{\lambda}^j}U_i,
\ee
$$
U_i=(U_{i1}, U_{i2}),
$$
and the coordinates $(x^1_i, x^2_i)$ with $i=1, 2, 3$ are the momentum maps of the triholomorphic vector
fields $\partial/\partial \theta$ and $\partial/\partial \varphi$.}
\\

  The Gibbons-Hawking form (\ref{gengibbhawk}) is the most appropriated to discuss supergravity solutions and for this
reason it will be instructive to check that (\ref{explico}) can be reduced to the form (\ref{gengibbhawk}).
But it is convenient first to explain  why
$(x^1_i, x^2_i)$ are the momentum maps of the isometries. In general momentum maps are related to a
compact Lie group $G$ acting over an hyperkahler manifold $M$ by triholomorphic Killing vectors
$X$, i.e, satisfying
$$
\pounds_X J^k=0.
$$
The last condition implies that $X$ preserves the Kahler-form $\Phi_k$, that is
$$
\pounds_X \Phi_k=0= i_X d\Phi_k+ d(i_X \Phi_k).
$$
Here $i_X \Phi_k$ denotes the contraction of $X$ with the Kahler forms.
By supposition $M$ is hyperkahler, then $d\Phi_k=0$ and
$$
d(i_X \Phi_k)=0.
$$
This implies that $i_X \Phi_k$ are a differential. The momentum maps $x^X_k$ are defined by
\be\lb{momap}
dx^X_k=i_X \Phi_k.
\ee
In the Pedersen-Poon case the isometries are $\partial/\partial \theta$ and $\partial/\partial \varphi$
and the hyperkahler form corresponding to (\ref{gengibbhawk}) is \cite{Gibbin}
$$
\Phi_k=(d\t_i+A_i)dx^i_k-U_{ij}(dx^i \wedge dx^j)_k
$$
where it should be identified $t_1=\theta$ and $t_2=\varphi$. From the last expression it follows that
$$
dx^{\theta}_k=i_{\theta}\Phi_k,\;\;\;\;dx^{\phi}_k=i_{\phi}\Phi_k
$$
and therefore $(x^1_i, x^2_i)$ are the momentum maps of the isometries.

     With this fact in mind it is possible to find the momentum map system $(x^i_{\theta}, x^i_{\varphi})$ for
(\ref{explico}). The contraction of $\partial/\partial\theta$ with the hyperkahler form (\ref{quakal}) gives
$$
dx_{\theta}^{1}=\frac{1}{\sqrt{\rho}F}(2q_0dq_2+2q_2dq_0-2q_1dq_3-2q_3dq_1
-(\frac{1}{2\rho}+\frac{F_{\rho}}{F})d\rho-\frac{F_{\eta}}{F}d\eta),
$$
$$
dx_{\theta}^{2}=\frac{1}{\sqrt{\rho}F}(2q_2dq_3+2q_3dq_2-2q_0dq_1-2q_1dq_0
-(\frac{1}{2\rho}+\frac{F_{\rho}}{F})d\rho-\frac{F_{\eta}}{F}d\eta),
$$
$$
dx_{\theta}^{3}=\frac{1}{\sqrt{\rho}F}(2q_0dq_0+2q_1dq_1-2q_2dq_2+2q_3dq_3
-(\frac{1}{2\rho}+\frac{F_{\rho}}{F})d\rho-\frac{F_{\eta}}{F}d\eta).
$$
The last expressions can be integrated to obtain
\be\lb{momentumap}
x_{\theta}^{1}=\frac{2(q_{0}q_{2} + q_{1}q_{3})}{\sqrt{\rho}F},\;\;\;
x_{\theta}^{2}=\frac{2(q_{2}q_{3} - q_{0}q_{1})}{\sqrt{\rho}F},\;\;\;
x_{\theta}^{3}=\frac{q_{0}^2 - q_{1}^2 - q_{2}^2 + q_{3}^2}{\sqrt{\rho}F}.
\ee
Similarly for $\partial/\partial \varphi$ it is found
$$
x_{\varphi}^{1}=\eta x_{\theta}^{1}+\frac{2\sqrt{\rho}( q_{1}q_{2}-q_{0}q_{3} )}{F},\;\;\;\;\;
x_{\varphi}^{2}=\eta x_{\theta}^{2}+\frac{\sqrt{\rho}(q_{0}^2 - q_{1}^2 + q_{2}^2 - q_{3}^2)}{F},
$$
\be\lb{momentumap2}
x_{\varphi}^{3}=\eta x_{\theta}^{3}+\frac{2\sqrt{\rho}(q_{0}q_{1} + q_{2}q_{3})}{F},
\ee
in accordance with \cite{Pedersen}.

     The next step is to determine the matrix $U_{ij}$ for (\ref{explico}). This is easily found
noticing that from (\ref{gengibbhawk}) it follows that
\be\lb{conditions}
U^{ij}=\overline{g}(\frac{\partial}{\partial t^i}, \frac{\partial}{\partial t^j}),
\;\;\;U^{ij}A_{j}=\overline{g}(\frac{\partial}{\partial t^i},\cdot).
\ee
  Then introducing the expression for (\ref{explico}) into the first (\ref{conditions}) gives
\be\lb{solmonop}
U^{ij}=\frac{|q|^2}{F(\frac{1}{4}F^2-\rho^2 (F_{\rho}^2+F_{\eta}^2))}\left(\begin{array}{cc}
  \frac{1}{2}F-\rho F_{\rho} & -\rho F_{\eta} \\
  -\rho F_{\eta} & \frac{1}{2}F+\rho F_{\rho}
\end{array}\right),
\ee
with inverse
\be\lb{impi}
U_{ij}=\frac{F}{|q|^2}\left(\begin{array}{cc}
  \frac{1}{2}F+\rho F_{\rho} & \rho F_{\eta} \\
  \rho F_{\eta} & \frac{1}{2}F-\rho F_{\rho}
\end{array}\right).
\ee
To find $A_i$ one should obtain from (\ref{Swannmetric}) and (\ref{metric}) that
$$
\overline{g}(\frac{\partial}{\partial \theta},\cdot)=\frac{1}{F}[(q_2\sqrt{\rho}
+ q_3\frac{\eta}{\sqrt{\rho}})(2q_1 \widetilde{A}^1-dq_0)
+ 2(q_2\frac{\eta}{\sqrt{\rho}}+ q_3\sqrt{\rho})(dq_1 + q_0 \widetilde{A}^1)
- 2(q_0\sqrt{\rho}+q_1\frac{\eta}{\sqrt{\rho}})(dq_2 + q_3 \widetilde{A}^1)
$$
$$
+ 2( q_0\frac{\eta}{\sqrt{\rho}}-q_1\sqrt{\rho})(dq_3-q_2 \widetilde{A}^1)],
$$
$$
\overline{g}(\frac{\partial}{\partial\varphi},\cdot)=\frac{1}{F}[\frac{q_3}{\sqrt{\rho}}(2q_1 \widetilde{A}^1-dq_0)
+ \frac{2q_2}{\sqrt{\rho}}(dq_1 + q_0 \widetilde{A}^1)
- \frac{2q_1}{\sqrt{\rho}}(dq_2 + q_3 \widetilde{A}^1)
+ \frac{2q_0}{\sqrt{\rho}}(dq_3-q_2 \widetilde{A}^1)],
$$
where $\widetilde{A}^1$ is given in (\ref{threefor}) in terms of F. Then from the second (\ref{conditions})
it is obtained
\be\lb{unoformas}
A_1=\frac{F}{|q|^2}[ (\frac{F}{2}+\rho F_{\rho}) \overline{g}(\frac{\partial}{\partial \theta},\cdot)
+\rho F_{\eta} \overline{g}(\frac{\partial}{\partial\varphi},\cdot) ],
\ee
\be\lb{unoformas2}
A_2=\frac{F}{|q|^2}[ \rho F_{\eta}\overline{g}(\frac{\partial}{\partial \theta},\cdot)
+ (\frac{F}{2}-\rho F_{\rho})\overline{g}(\frac{\partial}{\partial\varphi},\cdot) ].
\ee

    Therefore we have reduced (\ref{explico}) with this data to the form (\ref{gengibbhawk}).
Formulas (\ref{impi}), (\ref{unoformas}) and (\ref{unoformas2}) define
a class of solutions of the Pedersen Poon equations (\ref{genmonop})
and an hyperkahler metric (\ref{gengibbhawk}) simply described in terms of an unknown function
$F$ satisfying (\ref{backly}). We must recall however that this simplicity is just apparent because
$U_{ij}$ depends explicitly on the coordinates $(\rho, \eta, |q|^2)$, which depends implicitly
on the momentum maps $(x_{\varphi}^i, x_{\theta}^i)$  by (\ref{momentumap}) and (\ref{momentumap2}). Therefore
$U_{ij}$ is given only as an implicit function of the momentum maps.

     For physical applications it is important to find solutions which in this limit tends to
\be\lb{infinito}
\overline{g}= U^{\infty}_{ij}dx^i\cdot dx^j + U^{ij}_{\infty}dt_i dt_j,
\ee
for a constant invertible matrix $U^{\infty}_{ij}$ \cite{Gibbin}. Formulas (\ref{momentumap}) and (\ref{momentumap2})
shows that the asymptotic limit $x_{\theta}\rightarrow \infty$ or
$x_{\varphi}\rightarrow \infty$ corresponds to $q\rightarrow \infty$ or $\sqrt{\rho} F \rightarrow 0$.
In consequence from (\ref{impi}) and (\ref{solmonop}) it follows that $U^{ij}\rightarrow \infty$
and $U_{ij}\rightarrow 0$ asymptotically, which is not the desired result. This problem can be evaded defining
a new metric (\ref{gengibbhawk}) with
\be\lb{evade}
\overline{U}_{ij}=U_{ij} + U_{ij}^{\infty},
\ee
and with the same one-forms (\ref{unoformas}) and (\ref{unoformas2}) and coordinate system
(\ref{momentumap}) and (\ref{momentumap2}). Clearly to add this constant do not affect the solution
and this data is again a solution of the Pedersen-Poon equation
(\ref{genmonop}) for which $\overline{U}_{ij}\rightarrow U_{ij}^{\infty}$ and
$\overline{U}^{ij}\rightarrow U^{ij}_{\infty}$. Explicitly we have
\be\lb{evade2}
\overline{U}_{ij}=\frac{F}{|q|^2}\left(\begin{array}{cc}
  \frac{1}{2}F+\rho F_{\rho}+\frac{U_{11}^{\infty}|q|^2}{F} & \rho F_{\eta}+\frac{U_{12}^{\infty}|q|^2}{F} \\
  \rho F_{\eta}+\frac{U_{12}^{\infty}|q|^2}{F} & \frac{1}{2}F-\rho F_{\rho}+\frac{U_{22}^{\infty}|q|^2}{F}
\end{array}\right),
\ee
with inverse
\be\lb{evade3}
\overline{U}^{ij}=\frac{1}{det(\overline{U}_{ij})}\left(\begin{array}{cc}
  \frac{1}{2}F-\rho F_{\rho}+\frac{U_{22}^{\infty}|q|^2}{F} & -\rho F_{\eta}-\frac{U_{12}^{\infty}|q|^2}{F} \\
  -\rho F_{\eta}-\frac{U_{12}^{\infty}|q|^2}{F} & \frac{1}{2}F+\rho F_{\rho}+\frac{U_{11}^{\infty}|q|^2}{F}
\end{array}\right).
\ee

This modified metric is more suitable for physical purposes, but do not correspond to a Calderbank-Pedersen base
and this show that the converse of the Swann theorem is not necessarily true.

\section{Supergravity solutions related to hyperkahler manifolds}

   The hyperkahler spaces defined by (\ref{momentumap}), (\ref{momentumap2}), (\ref{impi}),
(\ref{unoformas}) and (\ref{unoformas2}) can be extended to 11-dimensional supergravity
solutions and to IIA and IIB backgrounds by use of dualities. This section present
them following mainly \cite{Gibbin}, more details can be found there and in
references therein.

   The hyperkahler solutions obtained in the previous section can be lifted
to D=11 supergravity solutions with vanishing fermion fields and $F_{\mu\nu\alpha\beta}$.
Such solutions are of the form
\be\lb{11sugra}
ds^2=ds^2(E^{2,1})+U_{ij}dx^i\cdot dx^j+ U^{ij}(dt_i+A_i)(dt_j+A_j),
\ee
and admits the action of a torus. Because the fields are invariant under the action
of the Killing vectors a solution of the IIA supergravity can be found
by reduction along one of the isometries, say $\partial/\partial\varphi$. The Kaluza Klein
anzatz is
\be\lb{KKanzatz}
ds^2=e^{-\frac{2}{3}\varphi(x)}g_{\mu\nu}(x)dx^{\mu}dx^{\nu}+e^{\frac{4}{3}\varphi(x)}
(dy + C_{\mu}(x)dx^{\mu})^2 ,
\ee
\be\lb{theform}
A_{11}=A(x)+ B(x)\wedge dy .
\ee
The field $A_{11}$ is the 3-form potential and $x^{\mu}$ are the coordinates of the D=10 spacetime.
The $NS \otimes NS$ sector is $(\phi, g_{\mu\nu}, B_{\mu\nu})$ and the
$R \otimes R$ sector is $(C_{\mu}, A_{\mu\nu\rho})$.  After reduction the nonvanishing fields
are
\be\lb{10sugra}
ds^2_{10}= (\frac{U_{11}}{detU})^2[ds^2(E^{2,1})+U_{ij}dx^i\cdot dx^j]
+(\frac{1}{U_{11}detU})^{1/2}(d\theta + A_{1})^2 ,
\ee
\be\lb{dilaton}
\phi=\frac{3}{4}log (U_{11})-\frac{3}{4}log (det U) ,
\ee
\be\lb{C}
C=A_2-\frac{U_{12}}{U_{11}}(d\theta + A_{1}).
\ee
All the quantities were independent of $\varphi$ and survived as Killing spinors of the
reduced theory.

   The field $\phi$ is independent of $\theta$ and $C$ satisfies $\pounds_{k} C=0$ and
one can use T-duality rules to construct a IIB supergravity solution
\be\lb{Tdual}
ds^2= [g_{mn}-g_{\theta\theta}^{-1}(g_{m\theta}g_{n\theta}-B_{m\theta}B_{n\theta})]dx^m dx^n
+2g_{\theta\theta}^{-1}B_{\theta n}d\theta dx^{n}+g^{-1}_{\theta\theta}d\theta^2
\ee
\be\lb{noseque}
\widetilde{B}=\frac{1}{2}dx^m \wedge dx^n [B_{mn}+ 2g_{\theta\theta}^{-1}(g_{m\theta}B_{n\theta})]
+ g_{\theta\theta}^{-1}g_{\theta m}d\theta \wedge dx^m
\ee
\be\lb{modilat}
\widetilde{\phi}=\phi-log(g_{\theta\theta})
\ee
where the tilde indicates the transformed fields. The restrictions
$$
B=0,\;\;\; i_{k}A=0,
$$
gives the T-dual fields
$$
\textit{l}=C_{\theta}
$$
\be\lb{restrict}
B'=[C_{mn}- (g_{\theta\theta})^{-1}C_{\theta}g_{\theta m}]dx^{m} \wedge d\theta ,
\ee
$$
i_{k}D=A ,
$$
where $\textit{l}$ is the IIB pseudoscalar, B' is the Ramond-Ramond 2-form
potential and D is the IIB 4-form potential. The non vanishing IIB fields
resulting from the application of the T-duality are
\be\lb{obtained}
ds^2_{10}= (detU)^{3/4}[(detU)^{-1}ds^2(E^{2,1})+(detU)^{-1} U_{ij}dx^i\cdot dx^j + d\theta^2] ,
\ee
\be\lb{B}
B_i=A_i\wedge d\theta ,
\ee
\be\lb{tau}
\tau=-\frac{U_{12}}{U_{11}}+i\frac{\sqrt{det U}}{U_{11}} ,
\ee
where
$$
\tau=\textit{l}+i e^{-\phi_B},\;\;\; B_{1}=B,\;\;\; B_{2}=B' ,
$$
and $ds^2_{10}$ is the Einstein frame metric satisfying
$$
ds^2_{10}=e^{-\phi_B/2}ds^2_{IIB} .
$$
More examples can be obtained by reducing (\ref{11sugra}) along one of the space directions
$E^{2,1}$ and it is obtained the IIA solution
\be\lb{flatred}
ds^2=ds^2(E^{1,1})+U_{ij}dx^i\cdot dx^j+ U^{ij}(dt_i+A_i)(dt_j+A_j),
\ee
with the other fields equal to zero. After T-dualizing in both angular directions it is obtained
$$
ds^2=ds^2(E^{1,1})+U_{ij}dX^i\cdot dX^j ,
$$
\be\lb{termi}
B=A_i \wedge dt_i,
\ee
$$
\phi=\frac{1}{2}log (det U) ,
$$
where $X^i={x^i, t^i}$. This solution can be lifted to a D=11 supergravity solution
\be\lb{extun}
ds^2_{11}= (detU)^{2/3}[(detU)^{-1}ds^2(E^{1,1})+(detU)^{-1} U_{ij}dX^i\cdot dX^j + d\theta^2],
\ee
\be\lb{extun2}
F=F_i\wedge dt_i \wedge d\theta.
\ee
  It is possible to generalize (\ref{11sugra}) to include a non vanishing 4-form $F$. The result
is the membrane solution
\be\lb{11sugra2}
ds^2=H^{-2/3}ds^2(E^{2,1})+H^{1/3}[U_{ij}dx^i\cdot dx^j+ U^{ij}(dt_i+A_i)(dt_j+A_j)],
\ee
\be\lb{11sugraF}
F=\pm\omega(E^{2,1})\wedge dH^{-1},
\ee
where $H$ is an harmonic function on the hyperkahler manifold, i.e, satisfies
$$
U^{ij}\partial_i \cdot \partial_j H=0.
$$
We have seen in the last section that every entry of $U_{ij}$ is an harmonic function and so such $H$ can be generated with
an hyperbolic eigenfunction F. After reduction along $\varphi$ it is found the following IIB solution
\be\lb{obtained2}
ds^2_{10}= (detU)^{3/4}H^{1/2}[H^{-1}(detU)^{-1}ds^2(E^{2,1})+(detU)^{-1} U_{ij}dx^i\cdot dx^j +H^{-1} d\theta^2],
\ee
\be\lb{B2}
B_i=A_i\wedge d\theta,
\ee
\be\lb{tau2}
\tau=-\frac{U_{12}}{U_{11}}+i\frac{\sqrt{det U}}{U_{11}},
\ee
\be\lb{queseyo}
i_{k}D=\pm\omega(E^{2,1})\wedge dH^{-1}.
\ee
If instead (\ref{11sugra2}) is dimensionally reduced along a flat direction it is obtained the IIA solution
\be\lb{11sugra3}
ds^2=ds^2(E^{1,1})+U_{ij}dx^i\cdot dx^j+ U^{ij}(dt_i+A_i)(dt_j+A_j),
\ee
\be\lb{sugra3}
B=\omega(E^{1,1})H^{-1},
\ee
\be\lb{dilaton}
\phi=-\frac{1}{2}log (H).
\ee
A double dualization gives a new IIA solution
\be\lb{dualizo}
ds^2=H^{-1}ds^2(E^{1,1})+U_{ij}dX^i\cdot dX^j,
\ee
\be\lb{ah}
B_i=A_i\wedge d\varphi^i+\omega(E^{1,1})H^{-1},
\ee
\be\lb{dilatame}
\phi=\frac{1}{2}log (det U)-\frac{1}{2}log (H).
\ee
The lifting to eleven dimensions gives
\be\lb{1extun}
ds^2_{11}= H^{1/3}(detU)^{2/3}[H^{-1}(detU)^{-1}ds^2(E^{1,1})+(detU)^{-1} U_{ij}dX^i\cdot dX^j + H^{-1}d\theta^2],
\ee
\be\lb{1extun2}
F=(F_i\wedge dt_i + \omega(E^{1,1})\wedge dH^{-1})\wedge d\theta.
\ee
All the backgrounds presented in this section can be constructed with a single F
satisfying (\ref{backly}), but the dependence on $(x_{\theta}, x_{\varphi})$ remains
implicit.

\section{Explicit and implicit solutions}
    In this section some particular solutions of the linear equations for F and V will be constructed
together with their Toda counterparts.

  The Ward equation $V_{\eta\eta}+\rho^{-1}(\rho V_{\rho})_{\rho}=0$
can be solved by separation of variables. The solutions obtained in this way are
$$
V_{0}=(A + B \eta)log (\rho),\;\;\;V_{+}=(C cos(\omega\eta)+ D sin(\omega\eta))(E I_{0}(\omega\rho)+ F K_{0}(\omega\rho)),
$$
\be\lb{solu1}
V_{-}=(G cosh(\omega\eta)+ H sinh(\omega\eta))(M J_{0}(\omega\rho)+ N Y_{0}(\omega\rho)),
\ee
where $\omega,A,B,..,N$ are constants, $J_0(\omega\rho)$ and $Y_0(\omega\rho)$ are Bessel functions
of first and second kind (or Newmann functions) and $I_0(\omega\rho)$ and $K_0(\omega\rho)$ are modified Bessel
functions  of first and second  kind (or MacDonald functions). From (\ref{shu}) it follows that
$$
G_\rho=-\rho V_{\eta};\;\;\;\;\;\;\ G_{\eta}=\rho V_{\rho}
$$
and $F=G/\sqrt{\rho}$ define separated solutions $F$ of (\ref{backly}) given by
$$
F_0=(A\rho^{3/2} + B \rho^{-1/2})(C\eta + D),
$$
\be\lb{solu2}
F_{+}=\rho^{1/2}(E I_1(\omega\rho)+ F K_1(\omega\rho))(G sin(\omega\eta)+ H cos(\omega\eta)),
\ee
$$
F_{-}=\rho^{1/2}(N J_1(\omega\rho)+ M Y_1(\omega\rho))(V sinh(\omega\eta)+ U cosh(\omega\eta)).
$$
This type of solutions has been used in \cite{yo} to construct certain $G_2$ holonomy examples.
One way to obtain non factorized solutions is to take the continuum limit of (\ref{solu1}) and (\ref{solu2})
by selecting $A,....,U,V$ as functions of $\omega$ and integrating in terms of this variable.

   The task of finding non separated solutions can be achieved selecting an arbitrary complex
function $H(z)$ in (\ref{intexp}) and solving (\ref{shu}). The powers $H(z)=z^n$
can be integrated out giving polynomial solutions. For instance selecting $H(z)=z^3$ gives
$$
V=3\eta \rho^2-2\eta^3;\;\;\;\;F=\frac{3}{4}\rho^{3/2}(4\eta^2-\rho^2).
$$
and a Toda solution holds by defining the coordinate system $(x,z)$
$$
x=3\rho^2-6\eta^2,\;\;\;\;\; z=6\eta\rho^2
$$
and $u(x,t)=log(\rho^2)$. Even in this simple case the coordinates
$(\rho, \eta)$ are given implicitly by the relations
$$
\rho^2=\frac{z}{6\eta},\;\;\;\;\; 6\eta^3+2\eta x-z=0.
$$
For $H(z)=z^5$ it is obtained
$$
V=\eta^5-5\eta^3\rho^2+\frac{15\eta\rho^4}{8};\;\;\;\;F=\frac{15\eta^2\rho^{7/2}}{4}-\frac{5\eta^4\rho^{3/2}}{2}
-\frac{15\rho^{11/2}}{48}.
$$
The powers $H(z)=z^n Log(z)$ can be also integrated explicit to give more complicated expressions
including logarithmic terms. Simple separated solutions has been used to study loop corrections
to the universal dilaton supermultiplet for type IIA strings on a Calabi-Yau manifold
\cite{Antoniadis}.

    One important class of non separated solutions are the $m$-pole ones, investigated in \cite{Dancer} and
\cite{Pedersen}. They give rise to the toric quaternionic Kahler metrics that are complete, compact and admitting only
orbifold singularities. Therefore the $G_2$ holonomy manifolds constructed with them as basis
are appropriated to discuss the appearance of non abelian gauge groups and chiral matter by
M-theory compactifications \cite{Witten2}, \cite{Angelova} and \cite{Angelito}. The basic eigenfunctions F
of (\ref{backly}) from which this solutions are constructed are
\be\lb{solu}
F(\rho, \eta, y)=\frac{\sqrt{(\rho)^2+(\eta-y)^2}}{\sqrt{\rho}}
\ee
where the parameter $y$ takes arbitrary values. Using the Backlund transformation it is found
the basic monopole
\be\lb{soludos}
V(\eta, \rho, y)=-Log[\eta-y + \sqrt{\rho^2 + (\eta-y)^2}].
\ee
The $2$-pole solutions are given by
$$
F_1=\frac{1+\sqrt{\rho^2+\eta^2}}{\sqrt{\rho}};\;\;\; F_2=\frac{\sqrt{(\rho)^2+(\eta+1)^2}}{\sqrt{\rho}}-
\frac{\sqrt{(\rho)^2+(\eta-1)^2}}{\sqrt{\rho}},
$$
The first one gives rise to the spherical metric and the second to
the hyperbolic metric
$$
ds^2=(1-r_1^2-r_2^2)^{-2}(dr_1^2+dr^2_2 + r_1^2d\theta_1^2 + r_2^2d\theta_2^2).
$$
The relation between the coordinates $(r_1, r_2)$ and $(\rho, \eta)$ can be extracted from
the relation
$$
(r_1 + i r_2)^2=\frac{\eta-1+i\rho}{\eta+1+i\rho}.
$$
The general "$3$-pole" solution is
$$
F=\frac{1}{\sqrt{\rho}}+\frac{b+c/m}{2}\frac{\sqrt{\rho^2+(\eta+m)^2}}{\sqrt{\rho}}
+\frac{b-c/m}{2}\frac{\sqrt{\rho^2+(\eta-m)^2}}{\sqrt{\rho}}.
$$
By definition $-m^2=\pm 1$, which means that $m$ can be imaginary or real.
The corresponding solutions are denominated type I and type II respectively and
encode many well known examples like the Bergmann metric on $CH^2$, the Eguchi-Hanson metrics,
the Bianchi VIII metrics, the bi-axial Bianchi IX metric and the Fubbini-Study metric on
$CP^2$ \cite{Pedersen}. There are also included to some quaternionic Kahler extensions
of hyperkahler metrics with two centers and $U(1)\times U(1)$ isometry \cite{Valent}.
3-pole solutions have many physical applications. They have been considered in \cite{Behrndt} to
construct type IIA solutions that can be interpreted as intersecting 6-branes. Moreover in \cite{Behrndt2},
the N=2 gauged supergravity coupled to the universal hypermultiplet
with a quaternionic geometry corresponding to the 3-pole solution has been considered,
and it has been studied the possibility to obtain the de Sitter vacua.

     The general $m$-pole solution is of the form
$$
F(\rho, \eta)=\sum^{\infty}_{k=0}\frac{\sqrt{a^2_k\rho^2+(a_k\eta-b_k)^2}}{\sqrt{\rho}},
$$
for some real moduli $(a_k, b_k)$ and the duality group SL(2, R)
acts over them \cite{Pedersen}. The corresponding monopole $V$ is
$$
V(\rho,\eta)=-\sum^{\infty}_{k=0} Log[a_k\eta-b_k+ \sqrt{a^2_k\rho^2 + (a_k\eta-b_k)^2}]
$$
and it can be checked that it satisfies (\ref{Wardy}).

    Another solution with application to string theories is \cite{Ketov},\cite{Ketov2}
and \cite{Behrndt2}
\be\lb{Ketov}
F(\rho,\eta)\sim f_{3/2}(\tau,\overline{\tau})=\sum_{(p,n)\neq(0,0)}\frac{\tau_2^{3/2}}{|p+n\tau|^3}
\ee
where $\tau=\tau_1+i\tau_2=\eta+i\rho$ and $f_{3/2}(\tau,\overline{\tau})$ is defined
by the Eisenstein series. Solution (\ref{Ketov}) is invariant under
the $SL(2,Z)$ duality
$$
\tau\rightarrow \frac{a\tau+b}{c\tau+d},\;\;\;ad-bc=1,\;\;\;(a,b,c,d)\in Z
$$
and describes the D-instanton corrections of the Universal Hypermultiplet moduli space preserving
some $U(1)\times U(1)$ symmetry, which are originated by Calabi-Yau wrapped two branes.

    Every solution $V$ presented here define implicitly a solution of the continuum Toda equation
by Proposition 5 and every $F$ define completely a toric quaternionic Kahler metric by proposition 3,
a toric 8 dimensional hyperkahler metric by (\ref{momentumap}), (\ref{momentumap2}),
(\ref{impi}), (\ref{unoformas}), (\ref{unoformas2}) and (\ref{gengibbhawk}), and different
supergravities solutions by the results of section 6.

\section{Conclusions}

  In the present work some important facts about toric quaternionic Kahler geometries in D=4
and toric hyperkahler geometries in D=8 has been presented and in particular,
it has been shown that the Swann construction provides a valid link between them. The fundamental
reason is that if a quaternionic Kahler space with torus isometry is used as a base in this
construction, then the two commuting isometries are preserved and are triholomorphic. As a result
the Swann extension of a toric quaternionic kahler space in D=4 is an
hyperkahler one with $T^2$ symmetry in D=8.
Because toric quaternionic spaces in D=4 are described in terms of a
simple linear equation (the Calderbank-Pedersen theorem), the resulting eight dimensional spaces
are also described in this manner. This result can be compared
with the Pedersen-Poon theorem that states that for toric hyperkahler metrics there exists a
coordinate system for which they take the
generalized Gibbons-Hawking anzatz. This system is related to the momentum
maps of the isometries, and to solutions of a generalized monopole
equation. As a consequence of this comparison there are found solutions for the generalized Gibbons-Hawking
anzatz just in terms of solutions the Calderbank-Pedersen equation, as
was shown in section 6. The simplicity of this solutions is just apparent,
because the dependence of the Gibbons-Hawking metric in terms of the momentum maps is
given implicitly.

 It should be remarked that there is not reason to state, in principle,
that all the eightdimensional toric hyperkahler spaces arise by the construction presented here.
We have presented only a subfamily among them in this paper. Such spaces were lifted
to 11-dimensional and IIA and IIB supergravity solutions by use of dualities,
and all the description is made as before in terms of a single F but with
implicit dependence on the momentum maps.

  The Calderbank-Pedersen description can be viewed as a consequence of the
Joyce classification of selfdual structures in four dimensions, together
with a result due to Tod and Przanowski.
Section 2 was intended to give the most simple presentation of the Joyce spaces
as possible. The Calderbank-Pedersen metrics are the Einstein representatives among the
Joyce families and are the most general local form of a quaternionic Kahler metric in
D=4 with $U(1)\times U(1)$ isometry. The conditions
arising from the demand of $T^2$ isometry together with the selfduality of
the Weyl tensor are very restrictive and as a consequence, all description is achieved only
in terms of the Joyce system, which is linear. The Einstein condition reduces this system
to a linear eigenvalue problem, that is, to find certain eigenfunctions F of the laplacian of the two dimensional
hyperbolic plane.

   This results presented in this paper can have different applications.
For instance proposition 3 describes the most general
local form of a selfdual Einstein metric with one isometry,
in terms of the Toda equation. They are the most general
quaternionic Kahler ones with one Killing vector
and can be extended to $G_2$ holonomy metrics by use of the Bryant-Salamon
construction \cite{Salamon}, that are asymptotically a cone over a
weak Calabi-Yau manifold. Such 7-dimensional metrics can be lifted to 11-supergravity
backgrounds and preserves $1/8$ of the supersymmetries after compactification.
The presence of one Killing vector allows a type IIA interpretation after
Kaluza-Klein reduction along the isometry.

    It will be also useful to analyze which are the symmetries of
the solutions of the Pedersen-Poon equations presented in this work.
In the well known brane solutions \cite{Gibbin} the dependence
in terms of the momentum maps is explicit,
and the symmetries of this solution related to momentum map system
are directly seen. But the solutions presented here, although they have the same
asymptotic limit, have not such explicit form and an interesting problem is
to find their symmetries.

\section{Acknowledgments}

We were benefited with mail correspondence with D.Joyce that made us clear certain features about his
work. We give thanks to A.Isaev, T. Mohaupt and S.Ketov for discussions about the subject and
its applications and to D.Mladenov, B.Dimitrov, T.Pilling and A.Oskin for many useful conversations.
O.P.S thanks specially to Luis Masperi for his permanent and unconditional support during his
mandate at the CLAF (Centro Latinoamericano de Fisica).

\appendix

\section{Generalities about Einstein-Weyl structures}

   The Einstein-Weyl structures are a generalization of the ordinary Einstein equation for
which exists a twistor correspondence \cite{Hitchin}. The Einstein equations are generalized
in this picture in order to include invariance under coordinate rotations plus dilatations.
In this section some important facts about them are sketched following \cite{Hitchin} and
\cite{JonTod}.

   It is known that for a given space $\overline{W}$ endowed with a metric $g_{ab}$
the Levi-Civita connection $\nabla$ is uniquely defined by
\be\lb{Levi-Civita}
\nabla g=0,\;\;\;T(\nabla)=0
\ee
where $T(\nabla)$ is the torsion. A Weyl-structure is defined by the manifold $\overline{W}$ together with:
\\

(a) A class of conformal metrics $[g]$, whose elements are related
by the conformal rescaling (or gauge transformation)
\be\lb{congauge}
g_{ab}\rightarrow \Omega^2 g_{ab},
\ee
together with $SO(n)$-coordinate transformations. $\Omega^2$ is an smooth, positive
real function over $\overline{W}$.
\\

(b) A torsion-free connection $D_{a}$ which acts over a representative
$g_{ab}$ of the conformal class $[g]$ as
\be\lb{Weyldef}
D_{a} g_{bc}=\omega_{a}g_{bc},
\ee
for certain functions $\omega_{a}$ defining an one form $\omega$. Then it
is said that $D$ preserves $[g]$.
\\

    The conformal group in $n$-dimensions is $CO(n)=R_{+}\times SO(n)$ and includes rotations plus
general scale transformations. The structure $[g]$ is called $CO(n)$-structure over $\overline{W}$.
The connection $D$ is uniquely determined by (\ref{Weyldef}) in terms of $\omega$ and $g$.
This can be seen in a coordinate basis $\partial_k$ in
which the system (\ref{Weyldef}) takes the form
$$
g_{ab,c}=g_{lb}\Upsilon^{l}_{ac} + g_{al}\Upsilon^{l}_{bc}+\omega_{a}g_{bc},
$$
where the symbols $\Upsilon^i_{jk}$ denotes the connection coefficients of $D$, which are symmetric
in the lower indices by the torsionless condition. Thus a series of steps analogous to
those needed to determine the Levi-Civita connection shows that $\Upsilon^i_{jk}$ is
\be\lb{nablacon}
\Upsilon^j_{ik}=\Gamma^j_{ik}+\gamma^j_{ik}
\ee
where $\Gamma^j_{ik}$ are the Christofel symbols and the add $\gamma^j_{ik}$ is
\be\lb{add}
2\gamma^i_{jk}=(\delta^i_j \omega_k + \delta^i_k \omega_j + g_{jk} \omega^i).
\ee

The form $\omega$ is not invariant under (\ref{congauge}), its transformation law
can be obtained from (\ref{Weyldef}) and (\ref{add}) and is
\be\lb{congauge}
\omega_{a}\rightarrow \omega_{a}+ 2\partial_{a}log(\Omega).
\ee
It follows from (\ref{add}) and (\ref{congauge}) that the Levi-Civita of any $g$ of $[g]$
preserves the conformal structure.

    If for a Weyl-structure the symmetric part of the
Ricci tensor $\widetilde{R}_{(ij)}$ constructed with $D_{i}$ satisfies
\be\lb{Einweyl}
\widetilde{R}_{(ij)}=\Lambda g_{ij}
\ee
for certain $\Lambda$, then will be called Einstein-Weyl.
If in addition the antisymmetric part of
$\widetilde{R}_{[ij]}$ vanish there exists a gauge in which
(\ref{Einweyl}) reduces to the vacuum Einstein equation
with cosmological constant $\Lambda$. To see this it is needed to calculate
$$
\widetilde{R}=D_{X}D_{Y}-D_{Y}D_{X}-D_{[X,Y]}
$$
using the formula (\ref{nablacon}) for $D$. The relation $CO(4)$=$R_{+}\times SO(4)$
decompose $\widetilde{R}$ into a real component $R_0$ and into an $SO(4)$-component
$R$ that is equal to the curvature tensor constructed with $\nabla$. After contracting indices
it is obtained
$$
\widetilde{R}_{ij}=R_{ij}+\nabla_i\omega_j-\frac{1}{2}\nabla_i\omega_j-\frac{1}{4}\omega_i\omega_j
+ g_{ij}(\frac{1}{2}\nabla_k\omega_k+\frac{1}{4}\omega_k\omega^k),
$$
where $R_{ij}$ is the Ricci tensor found with $\nabla$.
The antisymmetric part is originated by the $R_0$ component and is determined in terms of $\omega$ as
\be\lb{antim}
\widetilde{R}_{[ij]}=\frac{3}{2}\nabla_{[i}\omega_{j]}.
\ee
If (\ref{antim}) is zero, then $\omega$ is the gradient of certain function $\Psi$ over $\overline{W}$. The conformal
rescaling (\ref{congauge}) with $\Omega=-e^{\Psi}$ set $\omega=0$.
This reduce the symmetric part
$$
\widetilde{R}_{(ij)}=R_{ij}-\frac{1}{2}\nabla_{(i}\omega_{j)}-\frac{1}{4}\omega_i\omega_j
$$
to $R_{ij}$ and (\ref{Einweyl}) is the Einstein equation with $\Lambda$, thus $[g]$
contains an Einstein metric.

    In 3 dimensions Einstein-Weyl structures are related to the solutions
of the continuum limit of the Toda equation. If the anzatz for the metric
\be\lb{conformal}
h=e^{u}(dx^2+dy^2)+dz^2,
\ee
is introduced in the Einstein equation (\ref{Einweyl}) then it follows that
$u$ should satisfy \cite{Ward}
\be\lb{Toda}
(e^u)_{zz} + u_{yy} + u_{xx}=0,
\ee
\be\lb{conformal2}
\omega=-u_{z}dz.
\ee
The non linear equation (\ref{Toda}) is known as the $SU(\infty)$ Toda equation, and
is seen (\ref{conformal2}) that $\omega$ is entirely determined in terms of $u$. This equation
is integrable, but not many explicit solutions are known.

\section{The Joyce description of selfdual structures}

    One way to construct selfdual structures in four dimensions is to use the definition,
i.e, to take an anzatz for a metric tensor $g$, find the Levi-Civita connection and solve
the system of equations corresponding to $W_{-}=0$, then $[g]$ will be selfdual.
By another side it seems more natural to describe an structure $[g]$ in terms of a connection
$D$ preserving it like (\ref{Weyldef}) than in terms of the Levi-Civita one. This was done by Joyce
who considered which conditions should satisfy the curvature and the torsion of a
connection $D$ preserving $[g]$ in order to insure selfduality \cite{Joyce}. Some
important results have been successfully reformulated in this context, but the novelty is that it gives rise
to the classification of all 4 dimensional self-dual structures with two commuting isometries
that are surface orthogonal.

     The first question that arise is if it is possible to express
$W$ in terms of $D$ in order to impose
$W_{-}=0$ as a condition on $D$. It is well known that the irreducible components
of the Riemann tensor under the action of $SO(4)$ are
the scalar curvature $S$, is the traceless part of the Ricci tensor $R_{ij}$ and
the two components $W_{\pm}$ of the Weyl tensor \cite{Thorpe}-\cite{AtiyahSin}.
But this result is valid only for the curvature constructed with the $SO(4)$ Levi-Civita connection
$\nabla$. If instead a torsionless $CO(4)$-connection
$D$ preserving $[g]$ is considered, the relation
$CO(4)=R_{+}\times SO(4)$ splits $D$ into a real valued connection and an
$SO(4)$ connection and therefore the corresponding curvature
$\widetilde{R}$ is divided into an $SO(4)$ component $R(D)$ and into
an $R$-component $R_0(D)$. The irreducible parts are in this case 6
and among them there are two, denoted as $W_{\pm}(D)$, which are equal
to $W_{\pm}$ (see for instance Appendix A of \cite{Nuro}). Therefore $W$ can be
described in terms of a torsionless $D$ preserving $[g]$.

    If $D$ is supposed to have torsion then $\widetilde{R}$ has $10$
irreducible parts \cite{Joyce} and $W_{\pm}(D)$ is different
from $W_{\pm}$, even in the limit $\omega\rightarrow 0$. Nevertheless,
a careful analysis shows that
if $T(D)$ is selfdual, then $W_{-}(D)=W_{-}$. From this discussion
holds the following important result \cite{Joyce}:
\\

{\bf Proposition 8} {\it If for a conformal class $[g]$ it exists a connection
preserving $D$ for which
\be\lb{Joycetheorem}
T_{-}(D)=W_{-}(D)=0,
\ee
then $[g]$ is selfdual.}
\\

Conditions (\ref{Joycetheorem}) shows that selfdual $[g]$
can be characterized in terms of a connection $D$ preserving it
if the torsion is selfdual. Proposition 8 is a powerful method
to construct selfdual families $[g]$, although it is not the most
general one.

   The Ashtekar-Jacobson-Smolin description of selfdual Einstein spaces
arise as a simple application of Proposition 8. Consider a manifold $M$ and
four vector fields $e^{a}$ forming an oriented basis for $TM$ at each point,
and the metric $g$ constructed with $e^a$. The parallelizing connection
$D$ satisfies $De^{a}=0$, in this basis the Christofel symbols
are all zero and so $R(D)=0$. For this reason
the condition $W^-(D)=0$ is trivially satisfied. $D$ preserves the metric $g$, and the class $[g]$ of $g$.
The torsion is $T(e^a,e^b)=D_{e^a}e^b-D_{e^b}e^a-[e^a, e^b]=-[e^a,e^b]$
and has the anti-selfdual components
$$
T(e^1, e^2)-T(e^3,e^4),\;\;\; T(e^1, e^3)-T(e^2,e^4),\;\;\; T(e^1, e^4)-T(e^2,e^3),
$$
thus $T^{-}(D)=0$ if and only if
\be\lb{ashtekar}
[e^1, e^2]-[e^3,e^4]=0,\;\;\; [e^1, e^3]-[e^2,e^4]=0,\;\;\; [e^1, e^4]-[e^2,e^3]=0.
\ee
By Proposition 8 the equations (\ref{ashtekar})
defines a selfdual structure $[g]$; it is known as the Ashtekar-Jacobson-Smolin formulation
of the selfdual Einstein equations \cite{Ashtekar}. In particular selecting
$e^j=f_j\partial/\partial x_1 + \partial/\partial x_j$ it is
obtained
$$
\frac{\partial f_1}{\partial x_2}-\frac{\partial f_3}{\partial x_4}+\frac{\partial f_4}{\partial x_3}=0,\;\;
\frac{\partial f_1}{\partial x_3}-\frac{\partial f_4}{\partial x_2}+\frac{\partial f_2}{\partial x_4}=0,\;\;
\frac{\partial f_1}{\partial x_4}-\frac{\partial f_2}{\partial x_3}+\frac{\partial f_3}{\partial x_2}=0.
$$
If the metric has the Killing vector $\partial/\partial x^1$, then the functions $f_{i}$
are independent of $x_1$ and this system reduces to
\be\lb{Gibb-Hawk}
\nabla U=\nabla \times \omega
\ee
where we have defined $U=f_1$ and $\omega=(f_2, f_3, f_4)$. The
corresponding metric is
\be\lb{ashgib}
g=V^{-1}(dt-\omega)^2 + V dx \cdot dx,
\ee

It has been proved that (\ref{ashgib}) describes all
four dimensional selfdual examples that are Ricci-flat
with a triholomorphic $U(1)$-isometry, they are
known as the Gibbons-Hawking metrics \cite{GibbHawko}.

   To finish it should be mentioned than the Jones-Tod correspondence and the Joyce description of
selfdual structures with $U(1)\times U(1)$ isometry arise as a consequence of proposition 8
but the task to find the connection $D$ is more difficult. The reader interested in details can look at
the original reference \cite{Joyce}.


\begin{thebibliography}{99}

\bibitem{Atiyo} M. Atiyah and N.J.Hitchin The geometry and dynamic of magnetic monopoles,
Princeton University Press 1988.
\bibitem{AHDM} M.Atiyah, V.Drinfeld, N.Hitchin and Y.Manin Construction of instantons Phys.Lett.A 65 (1978) 185.
\bibitem{Gaume} L.Alvarez-Gaume and D.Z.Freedman  Geometrical structure and ultraviolet finiteness
in the supersymmetric sigma model Comm.Math.Phys.80 (1981) 443.
\bibitem{Hitchon} N.J.Hitchin, A.Karlhede, U.Lindstrom and M.Rocek Hyperkahler metrics and supersymmetry
Comm.Math.Phys.108 (1987) 535.
\bibitem{Galicki} K.Galicki A generalization of the momentum mapping construction for
quaternionic Kahler manifolds Comm.Math.Phys.108 (1987) 117.
\bibitem{Bagger} E.Witten and J.Bagger Matter couplings in N=2 supergravity Nucl.Phys.B 222 (1983) 1.
\bibitem{Fre} P.Fre The complete form of N=2 supergravity and its place in the general
framework of $D = 4 N$ extended supergravities Nucl.Phys.Proc.Suppl.45BC (1996) 59.
\bibitem{Fre2} L. Andrianopoli, M. Bertolini, A. Ceresole, R. D'Auria,
S. Ferrara, P. Fre' and T. Magri N=2 supergravity and N=2 superyang-mills theory on general
scalar manifolds: sympletic covariance, gaugings and the momentum map J.Geom.Phys.23 (1997) 111.
\bibitem{Unge}  I.Y. Park and R. von Unge Hyperkahler quotients, mirror symmetry and
F theory JHEP 0003 (2000) 037; U.Lindstrom, M.Rocek and R.von Unge
Hyperkahler quotients and algebraic curves JHEP 0001 (2000) 022.
\bibitem{Behrndt} K. Behrndt and G. Dall'Agata Vacua of N=2 gauged supergravity derived from nonhomogenous
quaternionic spaces Nucl.Phys.B 627 (2002) 357.
\bibitem{Antoniadis} I.Antoniadis and B.Pioline Higgs branch, hyperkahler quotient and duality in susy
N=2 Yang-Mills theories Int.J.Mod.Phys. A12 (1997) 4907.
\bibitem{deWit} B.de Wit, M.Rocek, S.Vandoren Gauging isometries on hyperkahler cones and quaternion
Kahler manifolds Phys.Lett.B 511 (2001) 302.
\bibitem{Salamon} R. Bryant and S.Salamon On the construction of some complete metrics with exceptional holonomy
Duke Math. Journal 58 (1989) 829.
\bibitem{Witten2} B.Acharya and E.Witten Chiral Fermions from Manifolds of $G_2$ Holonomy, hep-th/0109152.
\bibitem{Rocco} B.de Wit, M. Rocek and S.Vandoren Hypermultiplets, hyperkahler cones and quaternion
Kahler geometry JHEP 0102 (2001) 039.
\bibitem{Berger} M.Berger Sur les groupes d'holonomie des varietes a connexion afine et des
varietes Riemanniennies Bull.Soc.Math.France.83 (1955) 279.
\bibitem{Wolf} A.Wolf Complex homogeneous contact manifolds and quaternionic symmetric spaces
J.Math.Mech.14 (1965) 1033; D.Alekseevskii Riemannian spaces with exceptional holonomy groups
Func.Anal.Appl.2 (1968) 11.
\bibitem{Ishihara} J.Ishihara Quaternion Kahlerian manifolds J.Diff.Geom.9 (1974) 483; S.Salamon
Quaternionic Kahler manifolds Invent.Math.67 (1982) 143.
\bibitem{Proeyen}  B.de Wit and A.Van Proeyen Special geometry, cubic polynomials and homogeneous
quaternionic spaces Commun.Math.Phys.149 (1992) 307.
\bibitem{Kronheimer} P.B.Kronheimer The construction of ALE spaces as hyperkahler quotients I
J.Diff.Geometry 29 (1989) 665; A Torelli type theorem for gravitational instanton
J.Diff.Geometry 29 (1989) 685.
\bibitem{Swann} A.Swann Hyperkahler and quaternionic Kahler geometry Math.Ann.289 (1991) 421.
\bibitem{Anguelova} L.Anguelova, M.Rocek and S.Vandoren Hyperkahler Cones and Orthogonal Wolf Spaces
JHEP 0205 (2002) 064.
\bibitem{mohap} L. Jarv, T. Mohaupt and F. Saueressig Effective Supergravity Actions for Flop Transitions,
 JHEP 0312 (2003) 047; M-theory cosmologies from singular Calabi-Yau compactifications, hep-th/0310174.
\bibitem{Gibbin}  J.P. Gauntlett, G.W. Gibbons, G. Papadopoulos and P.K. Townsend
Hyperkahler  manifolds and multiply intersecting branes Nucl.Phys. B500 (1997) 133.
\bibitem{Portugues} R. Portugues, Membrane solitons in eight-dimensional hyper-Kaehler backgrounds, hep-th/0311099.
\bibitem{Manton} G.Gibbons and N.Manton The moduli space of well separated BPS monopoles
Phys.Lett.B 356 (1995) 32.
\bibitem{Runche} G.W. Gibbons and P. Rychenkova  Hyperkahler quotient construction
of BPS monopole moduli spaces hep-th/9608085.
\bibitem{Dyon}  G. Papadopoulos and J. Gutowski The dynamics of D-three-brane dyons and toric hyperkahler
manifolds Nucl.Phys. B551 (1999) 650.
\bibitem{Caloron} T.Kraan Instantons, monopoles and toric hyperkahler manifolds Comm.Math.Phys. 212 (2000) 503.
\bibitem{Tong} J.Gauntlett, D.Tong and P.Townsend Supersymmetric intersecting domain walls in massive
hyperkahler sigma models Phys.Rev.D 63 (2001) 085001.
\bibitem{Tong2} J.Gauntlett, D.Tong and P.Townsend Multidomain walls in massive supersymmetric sigma model
Phys.Rev.D 64 (2001) 025010.
\bibitem{Poon} H.Pedersen and Y.S.Poon Hyperkahler metrics and a generalization of the Bogomolny
equations Comm.Math.Phys.117 (1988) 569.
\bibitem{Joyce}  D.Joyce Explicit construction of selfdual 4-manifolds Duke.Math.Journal 77 (1995) 519.
\bibitem{Pedersen} D.Calderbank and H.Pedersen, Selfdual Einstein metrics with torus symmetry, math.DG/0105263;
 D.Calderbank and M.Singer Einstein metrics and complex singularities, math.DG/0206229.
\bibitem{Przanowski} M.Przanowski Killing vector fields in selfdual, Euclidean Einstein spaces
with $\Lambda\neq 0$ J.Math.Phys.32 (1991) 1004.
\bibitem{Todo} K.P.Tod The $SU(\infty)$ Toda equation and special four dimensional metrics
in Geometry and Physics Aarhaus J.E.Andersen J.Dupont H.Pedersen and A.Swann eds. 1995;
Lecture Notes in Pure and Appl.Maths 184 Marcel Dekker New York 1997.
\bibitem{Hitchin} N.Hitchin Complex manifolds and Einstein's equations in  Twistor geometry and Non-linear
Systems H.D.Doebner and T.D.Palev eds. Springer 1982.
\bibitem{Ward} R.S.Ward  Einstein-Weyl spaces and $SU(\infty)$ Toda fields Class.Quant.Grav 7 (1990) L95;
N.Woodhouse Cylindrical gravitational waves Class.Quant.Grav 6 (1989) 933.
\bibitem{Thorpe} I.M.Singer and J.A.Thorpe The curvature of 4-dimensional Einstein spaces
in Global Analisis, Papers in honor of K.Kodaira D.C.Spencer and S.Iyanaga eds.
\bibitem{AtiyahSin} M.Atiyah, N.Hitchin and I.Singer Selfduality in four-dimensional riemannian geometry
Proc.R.Soc.London.A 362 (1978) 425.
\bibitem{Ashtekar} A.Ashtekar T.Jacobson and L.Smolin A new characterization of half flat solutions
to Einstein's equation Comm.Math.Phys.115 (1988) 631.
\bibitem{GibbHawko} G.Gibbons and S.Hawking Action integrals and partition functions in quantum
gravity Phys.Rev.D 15 (1977) 2752; K.P.Tod and R.S.Ward Selfdual metrics with selfdual Killing
vectors Proc.R.Soc.London.A 368 (1979) 411.
\bibitem{yo} O.P.Santillan A construction of $G_2$ holonomy spaces
with torus symmetry Nucl.Phys.B 660 (2003) 169.
\bibitem{Angelova} L.Anguelova and C.Lazaroiu, Enhanced gauge symmetry from `toric' $G_2$ cones, hep-th/0208177.
\bibitem{Angelito} L.Anguelova and C.Lazaroiu, M-theory compactifications on certain `toric' cones of $G_2$ holonomy
 hep-th/0204249.
\bibitem{Dancer} R. Bielawski and A.Dancer The geometry and topology of toric hyperkahler manifolds
Comm.Anal.Geom.8 (2000) 727.
\bibitem{Valent} P.Casteill, E.Ivanov and G.Valent $U(1)\times U(1)$ quaternionic metrics from
harmonic superspace Nucl.Phys. B627 (2002) 403.
\bibitem{Behrndt} K. Behrndt, G. Dall'Agata, D. Lust and S. Mahapatra
Intersecting 6-branes from new 7-manifolds with $G_2$ holonomy, JHEP 0208 (2002) 027.
\bibitem{Behrndt2} K. Behrndt and S. Mahapatra De Sitter vacua from N=2 gauged supergravity,  hep-th/0312063.
\bibitem{Ketov} S.V.Ketov D-instantons and Universal Hypermultiplet hep-th/0112012;
Summing up D-instantons in N=2 supergravity,Nucl.Phys. B649 (2003) 365.
\bibitem{Ketov2} S.V.Ketov Instanton-induced scalar potential for the universal hypermultiplet,
Nucl.Phys. B656 (2003) 63; D-instantons and matter hypermultiplet  Phys.Lett. B558 (2003) 119.
\bibitem{Antoniadis}  I. Antoniadis, R. Minasian, S. Theisen and P. Vanhove
String loop corrections to the universal hypermultiplet, Class.Quant.Grav. 20 (2003) 5079.
\bibitem{JonTod} P.E.Jones and K.P.Tod Minitwistor spaces and Einstein-Weyl spaces Class.Quant.Grav.2 (1985) 565.
\bibitem{Nuro} P.Nurowski Twistor bundles, Einstein equations and real structures Class.Quant.Grav 14 (1997) A261.
\end{thebibliography}
\end{document}